\documentclass[showpacs,amsmath,amssymb,twocolumn,floatfix,prl]{revtex4}
\usepackage[dvips]{graphicx}
\input{epsf}

\begin{document}

\title{Imperfection effects for multiple applications of 
the quantum wavelet transform}


\author{M. Terraneo and D. L. Shepelyansky}
\homepage[]{http://www.quantware.ups-tlse.fr}
\affiliation{Laboratoire de Physique Quantique, UMR 5626 du CNRS, 
Univ. Paul Sabatier, 31062 Toulouse Cedex 4, France}

\date{March 9, 2003}

\begin{abstract}
We study analytically and numerically the effects of various 
imperfections in a quantum computation of a simple dynamical
model based on  the
Quantum Wavelet Transform (QWT). 
The results for fidelity timescales, obtained for 
a large range of error amplitudes
and number of qubits, imply that for
static imperfections
the threshold for fault-tolerant
quantum computation is decreased by a few orders of magnitude  
compared to the case of random errors.
\end{abstract}

\maketitle

The mathematical theory of Wavelet Transforms (WT) finds nowadays an enormous 
success in various fields of science and technology, including 
treatment of large databases, data and image compression, signal processing,
telecommunications and many other applications \cite{Daub,Mey}. 
Wavelets  are obtained 
by translations and dilations of an original function and they allow to obtain
high resolutions of microscopic details, both in frequency and space.  
The discrete WT can be implemented 
with high computational 
efficiency and provide a powerful tool for treatment of digital data.
It is well accepted that the Fourier transform and WT are the main 
instruments for data treatment, and it has been shown that 
in many applications the performance of WT is  much higher 
compared to the Fourier analysis. 
The permanent growth of 
computer capacity has significantly
increased the importance of the above transformations in numerical 
applications. 

The recent development of quantum information processing 
has shown that 
computers based on laws of quantum mechanics can perform certain tasks 
exponentially faster than any known classical computational algorithms 
(see {\it e.g.} \cite{Chuang}). 
The most known example is the integer factorization algorithm 
proposed by Shor \cite{shor}. An essential element of this algorithm is 
the Quantum Fourier Transform (QFT) which can be performed for a vector of size
$N=2^{n_q}$ in $O(n_q^2)$ quantum gates, in contrast to $O(2^{n_q} n_q)$ 
classical operations \cite{Chuang,shor}. Here $n_q$ can be viewed as the number
of qubits (two-level quantum systems) of which a quantum computer is built.
 Apart from Shor's algorithm, the
QFT finds a number of various applications in quantum computation, 
including the simulation of quantum chaos models showing rich and complex 
dynamics \cite{shack,song,simone}. 
The sensitivity of the QFT to imperfections was tested in numerical 
simulations and the time-scales for reliable computation of the algorithm
were established \cite{cirac,paz,song,simone}. 

A few years after the discovery of the QFT algorithm, it has been shown that
certain WT can also be implemented on a quantum 
computer in a polynomial number of quantum gates \cite{hoyer,williams,klapp}. 
In fact, explicit quantum circuits were developed for the most popular
discrete WT, namely the 4-coefficient Daubechies WT
($D^{(4)}$) and the Haar WT, both for pyramidal and
 packet 
algorithms \cite{hoyer,williams,klapp}. 
As it happens in classical signal analysis, it is natural to expect that 
QWT will find important future applications for 
the treatment of quantum databases and quantum data compression. 
Therefore, it is important to investigate
the stability and the accuracy of QWT in respect to imperfections. 
This is especially important since the functions
of the wavelet basis have singularities in the derivatives 
(in contrast to analyticity of 
Fourier waves) that may enhance the effects of perturbations.
 
To this aim we introduce a simple model with rich nontrivial dynamics
which is  essentially based on multiple applications of the WT.
Its quantum evolution can be 
efficiently simulated on a quantum computer, and it is described by the 
unitary map for the wave function $\psi$:
\begin{equation}
\bar{\psi} = \hat{U} \psi = \hat{W}^{\dagger} e^{-i k (x-\pi)^2/2}  
\hat{W} e^{ -i T n^2/2} \psi . 
\label{model}
\end{equation}
\noindent
Here the bar marks the value of the wave function after one map iteration,
 $\hat{W}$ is the $D^{(4)}$ WT operator, and the unitary 
diagonal operators $U_T=e^{ -i T n^2/2}$ and $U_k=e^{-i k (x-\pi)^2/2}$
represent quantum phase rotation in computational and wavelet basis, 
respectively. The evolution takes place in the Hilbert space of $N=2^{n_q}$
states, with $-N/2 \le n < N/2$ and $x=2 \pi j/N $ where  $j=0,\ldots N-1$ is 
the index in the wavelet basis and $T$, $k$ are dimensionless parameters.
In the case when $\hat{W}$ is replaced by the Fourier transform, 
one obtains the quantum sawtooth map previously analyzed in Ref.\cite{simone}.
Thus the model (\ref{model}) can be considered as a 'kicked wavelet 
rotor', where $k$ is the kick strength in the wavelet basis. 
We numerically 
tested that the dynamical properties are not very sensitive to the value 
of $T$ and here we present data for a typical value $T=1.4$.

\begin{figure}[t!]  
\centerline{\epsfxsize=4.2cm\epsffile{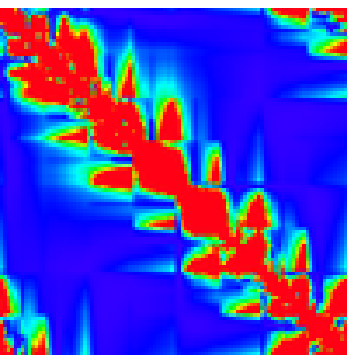}
\hfill\epsfxsize=4.2cm\epsffile{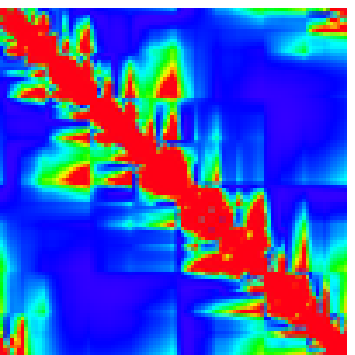}}
\centerline{\epsfxsize=4.2cm\epsffile{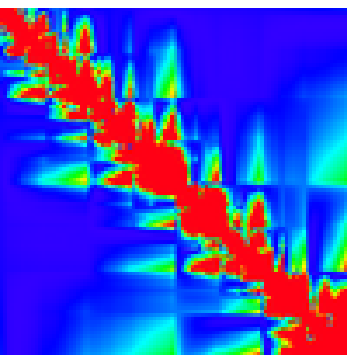}
\hfill\epsfxsize=4.2cm\epsffile{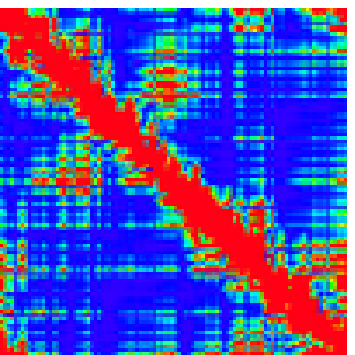}}
\vglue -0.3cm 
\caption{(Color on line) 
Density plot of matrix elements $|U_{n,n'}|^2$ for the model (\ref{model}) 
in the computational basis, for 
 $N=2^{12}$. Top:
$k=100$ (left), $k=1000$  (right); bottom is for $k=1000$:
a doubled resolution of left upper quarter (left),
perturbed operator with static errors $\epsilon =5 \times 10^{-4}, \mu=0$.
Color marks the density from blue (zero) to red (maximal value).
}
\label{fig1}       
\end{figure}

\begin{figure}[t!]
\epsfxsize=3.2in
\epsfysize=2.6in
\epsffile{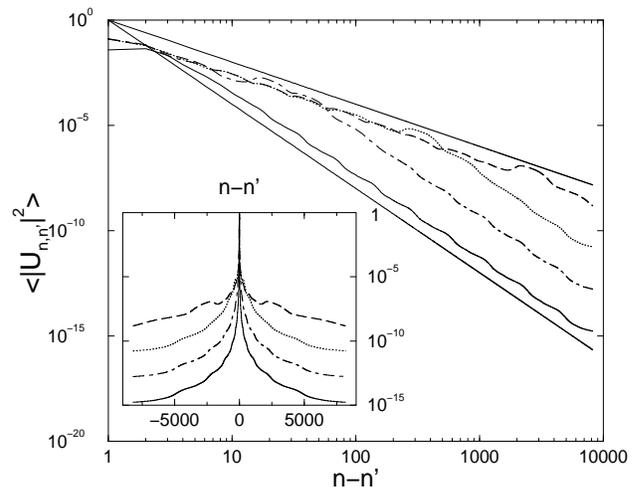}
\vglue -0.5cm
\caption{
Dependence of averaged matrix elements $\langle |U_{n,n'}|^2\rangle $ on
$|n-n'|$ (the average is taken along the diagonal). Data are shown for 
$N=2^{15}$ and $k=1$ (full black curve), $k=10$ (dash-dotted curve),
$k=100$, (dotted curve) and $k=1000$ (dashed curve).
The two straight lines are $1/|n-n'|^2$ and $1/|n-n'|^4$. 
The inset shows the data in semi-log scale.
}
\label{fig2}
\end{figure}

The global properties of the evolution operator (\ref{model}) are shown in 
Figs.\ref{fig1},\ref{fig2} for different values of $k$
 (see also Appendix A1). 
The density 
plot of transition matrix elements  $U_{n,n'}$ in the computational 
basis is 
represented in Fig.\ref{fig1}. By increasing $k$ a larger and larger number
of states is coupled by the dynamics, and the complex 
self-similar structure of the 
transitions generated by the WT becomes evident.
On the average, the off-diagonal matrix elements decay with the power law
$|U_{n,n'}|^2 \sim 1/|n-n'|^{\alpha}$. Asymptotically for 
$|n-n'| \gg 5k $ we obtain the exponent $\alpha=4$ (Fig.\ref{fig2}).
 For large values of $k$ the intermediate scaling law is described by the 
exponent $\alpha=2$, in the range $1\le |n-n'| \ll 5k$.
This decay law for the matrix elements can be considered as a long range 
coupling between states. We note that similar power law regimes have been 
analyzed in random matrix models \cite{bouch,fyo}.
Our numerical analysis shows that there are two regimes for the level 
spacing statistics $P(s)$ \cite{mirlin} in the limit of large $N$. {\it E.g.} 
for $N=2^{12}$ 
  the distribution $P(s)$ is given by the Poisson law for $k < 5$, while  for
$5 < k \le 10000$  it shows level repulsion and a poissonian 
decay for large $s$ (see Appendix A2).
We attribute the rapid appearance of level
repulsion to the slow power law decay of matrix elements \cite{fyo}.

To implement the evolution (\ref{model}) on a quantum computer, 
we developed an
 algorithm based on the QWT for the Daubeschies $D^{(4)}$ wavelets.
The algorithm consists of four steps: i) the multiplication by $\hat{U}_T$, 
performed in $O(n_q^2)$ controlled-phase shift gates as described in 
\cite{simone}; ii) the application of $\hat{W}$ operator,  realized by the 
QWT following
the circuit described in Fig.10 of \cite{williams} 
(see Appendix A3); 
iii) the operator 
$\hat{U}_k$, implemented in a similar way as for the step i); iv)the inverse 
WT $\hat{W}^{\dagger}$, obtained by reversing the gates
of the step ii).
The heaviest parts of the algorithm are the steps ii), iv), since the 
QWT algorithm requires multi-controlled operations. To implement them we used 
the recipe given in \cite{vedral} which allows to realize a $n$-controlled
gate by $O(n)$ elementary gates (Toffoli and 1- and 2-qubit gates). 
To this end an 
ancilla qubit is needed, so that we used $n_q+1$ qubits to 
simulate numerically the dynamics of model (\ref{model}) 
with $N=2^{n_q}$ states.
The implementation of the wavelet kernel $D_{2^n}^{(4)}$ requires $O(n)$ 
multi-controlled gates ($n=2,\ldots,n_q$), and since the QWT is composed 
of $O(n_q)$ kernel 
applications this leads the total number of elementary gates to scale as
 $O(n_q^3)$ \cite{hoyer,williams,klapp}
(see Appendix A3).
To study the algorithm accuracy we consider two models of imperfections.
In the model of random noisy gates we replace all ideal gates
by imperfect ones, which are obtained by  random unitary rotations
by a small angle $\eta$, $-\epsilon/2 \le \eta  \le \epsilon/2$, around 
the ideal rotation angle (as in \cite{cat}). 
In the model of static imperfections (see \cite{georgeot,simone})
all gates are perfect
but between gates $\psi$ accumulates a
phase factor  $e^{i \phi }$ 
with $\phi = \sum_l (\eta_l \sigma^z_l+
\mu_l\sigma^x_l \sigma^x_{l+1})$. Here $\eta_l, \mu_l$ vary randomly with
$l=0,...,n_q$, $\eta_l$ represents static  one-qubit energy shifts,
$-\epsilon/2 \le \eta_l  \le \epsilon/2$,
and $\mu_l$ represents static inter-qubit couplings on a circular chain,
$-\mu/2 \le \mu_l  \le \mu/2$.

\begin{figure}[t!]
\epsfxsize=3.2in
\epsffile{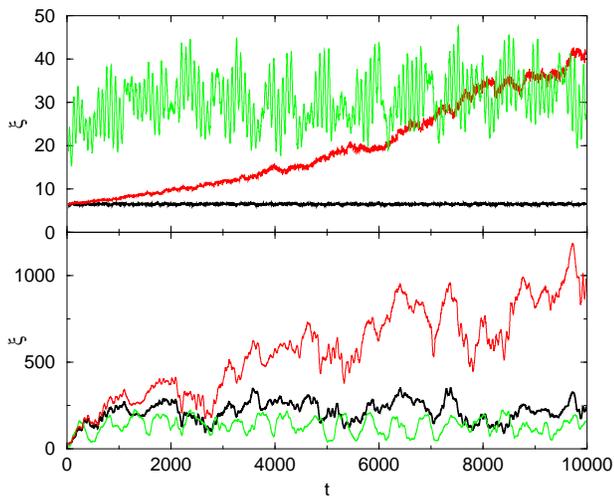}
\vglue -0.5cm
\caption{(Color on line) 
Dependence of IPR $\xi$ on the number of iterations $t$, for $n_q=12$,
 $T=1.4$,  $k=1$ (top) and $k=1000$ (bottom).
Initially the probability is concentrated at $n=0$.
The black curves show the quantum computation  with ideal gates; 
the green (light gray) curves show  the case with static errors 
at $\epsilon=10^{-4}, \mu=0$  and  
red (gray) curves correspond to the case with noisy gates
at $\epsilon=5 \times 10^{-4}$. The data are
averaged over time interval $\Delta t = 50$.
}
\label{fig3}       
\end{figure}

\begin{figure}[t!]
\epsfxsize=3.2in
\epsffile{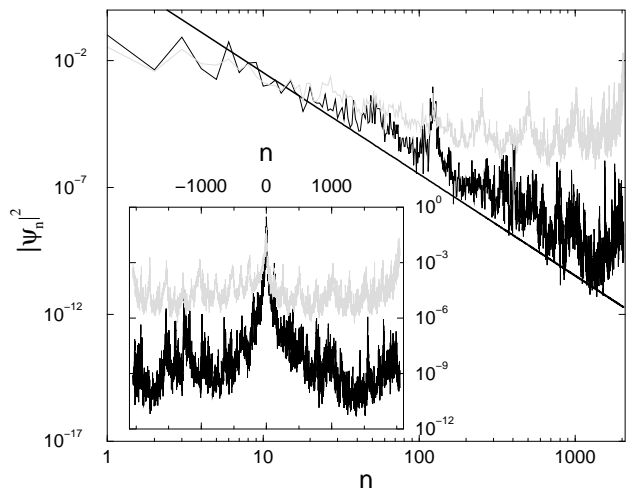}
\vglue -0.5cm
\caption{
Probability distribution $|\psi_n|^2$ in the computational basis
for the parameters of Fig.\ref{fig3} (top) at $k=1, t=10^4$:
full curve is the quantum computation with ideal gates, gray curve
shows data for noisy gates with $\epsilon=5 \times 10^{-4}$. The straight line 
displays the scaling law $1/n^4$.
The inset shows the same data in semilogarithmic scale.
}
\label{fig4}       
\end{figure}

The numerical simulations of the ideal quantum algorithm for the map 
($\ref{model}$) show that the wave function is essentially localized on a few 
states of the computational basis. This localization is clearly seen from 
the Inverse Participation Ratio (IPR) $\xi = 1/\sum_{n} |\psi_n|^4$ which 
is a standard quantity to characterize localization in mesoscopic  systems 
\cite{mirlin}. It directly provides the number of sites on which the 
probability is concentrated. Surprisingly the localization is present 
not only for moderate $k \sim 1$, but also when the kick strength is  very
large $k \sim 1000$ (see Fig. \ref{fig3} and Appendix A4).
Indeed in both cases $\xi$ 
fluctuates near a constant value $\xi_0 \ll N$, 
even for a very large number of iterations.
We attribute this localization to the structure of the operator 
(\ref{model}): it is banded for moderate $k$ and sparse for large $k$ (see 
Fig.(\ref{fig1})). For $k \sim 1$ the probability shows an 
algebraic localization $|\psi_n|^2 \propto 1/n^4$ (Fig.\ref{fig4}). 
Such an exponent fully agrees  with the scaling law of Fig.\ref{fig2}.
For $k>100$, the probability is spread over the whole basis (data not shown), 
but only a moderate number of narrow peaks contributes to the IPR value 
(see Fig.\ref{fig3}).
This behaviour is consistent with the fact that the $P(s)$ never reaches a 
Wigner-Dyson regime (see discussion above). On the contrary, the 
spectral properties of the sawtooth
map \cite{simone,simone1} are described by the 
random matrix theory for $k\sim 1000$, $T\sim 1$ and $N=2^{12}$.

The effect of imperfections in the quantum gates is shown in Figs. \ref{fig3},
\ref{fig4} (see also Appendix A4). 
The results clearly show that the localization is destroyed by noisy
gates imperfections which lead to an approximately linear growth 
of $\xi$ with $t$. For static imperfections $\xi$ shows modified
bounded oscillations.
The probability distribution in Fig.\ref{fig4} shows the appearance of a 
plateau with 
pronounced peaks located approximately at $n = N/2^m$, $m=1, 2, 3 \ldots$.
We attribute the appearance of these peaks to the pyramidal structure of the 
algorithm, which in the presence of imperfections produces stronger errors at 
the  above values of $n$. For static imperfections
the plateau level remains bounded in time $t$ while
for noisy gates it increases with $t$ and 
for very large $t$  the probability becomes homogeneously 
distributed over the computational basis.

\begin{figure}[t!]
\epsfxsize=3.2in
\epsffile{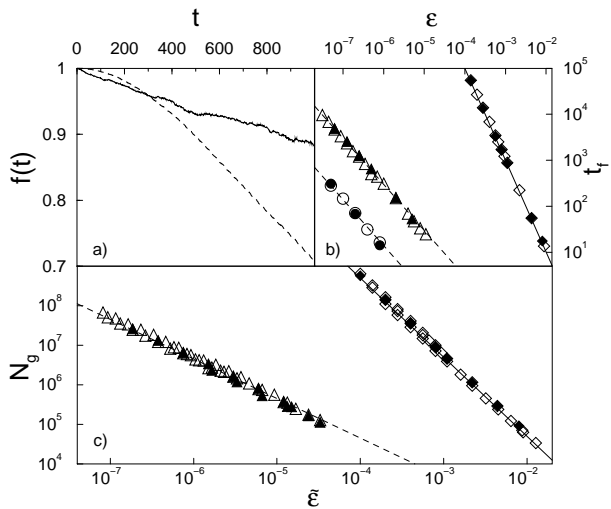}
\vglue -0.5cm
\caption{
Panel a) shows the fidelity decay in time at $k=1$, $n_q=12$ for static 
imperfections
($\epsilon=10^{-4}, \mu=0$, dashed curve) 
and noisy gates ($\epsilon = 5 \times 10^{-4}$, full curve).
Panel b) shows the dependence of time scale $t_f$ 
on the imperfection strength 
$\epsilon$ for $n_q=8$ ($n_g=5237$) for noisy gates (diamonds)
and static imperfections (triangles at $\mu=0$; circles at
$\mu=\epsilon$,  for clarity data are shifted in 
$\epsilon$-axis by factor 10
to the left).
Panel c) gives the dependence of the total number of gates $N_g$
on $\tilde{\epsilon} $ for $n_q=6,8,10$. For noisy gates (diamonds)
 $\tilde{\epsilon} =\epsilon$ and for static imperfections 
(triangles) $\tilde{\epsilon} =\epsilon \sqrt{n_q}$.
Open(full) symbols are data for $k=1(k=1000)$. The full and dashed lines 
in panels b)/c) show the relations 
(\ref{tf1}) and (\ref{tf2}), respectively.
}
\label{fig5}
\end{figure}

The qualitative  difference between two types
of imperfections becomes clear from the analysis of the behaviour of the 
fidelity, defined as $f(t)=|\langle \psi_{\epsilon}(t)|\psi(t)\rangle|^2$.
Here $\psi(t)$ is the wave function obtained with ideal gates, while
$\psi_{\epsilon}(t)$ is the result of the quantum computation with 
imperfections of amplitude $\epsilon$. 
We determine the time scale $t_f$ for accurate
computation by fixing a threshold for the fidelity as $f(t_f) = 0.9$. 
In this way it is possible to find the dependence of $t_f$ on the system 
parameters. Our numerical data are presented in Fig.\ref{fig5}.
They show that for noisy gates $t_f$ is described by the relation
\begin{equation}
t_f= C/(\epsilon^2 n_g), \;\;\; N_g = C/\epsilon^2
\label{tf1}
\end{equation}
where $n_g$ is the number of gates per map iteration,
$N_g=n_g t_f$ is the total number of gates and $C \approx 5$ is a 
numerical constant. The physical origin for this scaling is related to the
fact that after each gate an amount of probability of the order of 
$\epsilon^2$  is transferred from the ideal state to all other states. This 
leads to an exponential decay of the fidelity  $f(t) \approx
\exp(-A  \epsilon^2 n_g t)$, where $A$ is a constant (see Fig.\ref{fig5}a).
This  gives the scaling (\ref{tf1}), which was also found in other algorithms
with noisy gates \cite{paz,simone,cat}.

For the model with static imperfections the scaling is 
\begin{equation}
t_f=D/(\epsilon n_g {n_q}^{1/2}), \;\;\; N_g=D/(\epsilon {n_q}^{1/2})
\label{tf2}
\end{equation}
where $D$ is a numerical constant
($D \approx 4.5$, at $\mu=0$ and $D \approx 2.1$
at $\mu=\epsilon$). This timescale is significantly
 smaller than the one for noisy gates. Physically, this happens due to 
the coherent action of static imperfections, which lead to effective 
Rabi oscillations proportional to $\cos{(\epsilon  n_g t)}$ for each qubit. 
For $n_q$ qubits this gives $f(t) \propto [\cos{(\epsilon n_g t)}]^{n_q}$ and
for small $\epsilon$ we obtain a Gaussian drop of the fidelity
 $f(t) \sim \exp{(-n_q (\epsilon n_g t)^2)}$, in agreement with our 
numerical results (see Fig.\ref{fig5}a). 
This leads to the scaling (\ref{tf2}), which is
confirmed by the data in Fig.\ref{fig5}. The effects of static imperfections
are dominant for all range of imperfection strengths studied.
We note that the scaling laws
(\ref{tf1}), (\ref{tf2}) are rather general and do not depend on the kick
strength $k$.
We note that similar scalings 
 were discussed  and numerically demonstrated in other
quantum algorithms with noisy gates \cite{paz,cat} and static imperfections
\cite{simone} (see also \cite{zurek}).
This shows that such scaling laws are generic and are not
sensitive to the singularities in the derivatives of the wavelets.
The universality of the above relations (\ref{tf1}), (\ref{tf2}) is also 
confirmed by the fact that the structure of the QWT is rather different from 
the QFT algorithm, {\it e.g.} the number of elementary quantum gates scales 
as $O(n_q^3)$ for the QWT, in contrast to $O(n_q^2)$ for the QFT. 
These relations  determine the total number of gates $N_g
= t_f n_g$ during which the quantum computation is reliable.
Similar scalings for $N_g$ 
should also be valid for other quantum algorithms, {\it e.g.} Grover's and
Shor's algorithms. We discuss also other types of errors in 
Appendix A5.

The above relations (\ref{tf1}), (\ref{tf2}) are important for the 
quantum error correction codes and the fault-tolerant quantum computation
threshold (see \cite{Chuang,steane} and Refs. therein).
Indeed the accuracy border for large scale quantum computation
is obtained in the assumptions of random  noisy errors and gives a 
threshold $\epsilon < \epsilon_r \sim 10^{-2}$. This approach
intrinsically uses the fact that for noisy gates the fidelity remains 
close to one for a number of gates $N_g = C/\epsilon_r^2$ 
(see (\ref{tf1})). In the case of static imperfections it is natural 
to assume that this number of gates should remain approximately the same
to allow large scale computation on a quantum computer
with $n_q$ qubits. Therefore, for static imperfections the 
Eqs. (\ref{tf1}), (\ref{tf2}) give the accuracy border
$\epsilon_s$:
\begin{equation}
\epsilon_s \approx D \epsilon_r^2/(C {n_q}^{1/2})
\label{tf3}
\end{equation}
This important relation gives a significant decrease of the 
threshold for the case of static imperfections \cite{note1}.
For the parameters of our model at $n_q=10$ we obtain 
that for the noisy error rate
$p_r=\epsilon_r^2 \approx 10^{-4}$ 
the rate induced by  static imperfections should be less than
$p_s=\epsilon_{s}^2  \approx 10^{-9}$. 
This result shows that new strategies of 
quantum error correction codes should be developed to
significantly suppress  phase shifts induced by static imperfections.
The spin echo techniques used in NMR \cite{Chuang} may 
play here an important role.

\begin{acknowledgments}
This work was supported in part by the EC contracts RTN QTRANS and IST-FET
EDIQIP and the NSA and ARDA under ARO contract No. DAAD19-01-1-0553. We thank 
CalMiP in Toulouse and  IDRIS at Orsay for access to their supercomputers.
\end{acknowledgments}

\begin{widetext}

\newpage
\section{APPENDIX}
\section{A1} Here we show examples of density plot for the matrix elements
$|U_{n,n'}|^2$ for the model (\ref{model}) for $k=1, 10, 100, 1000$
(Fig.\ref{Afig1}).
\begin{figure}[h!]
\begin{center}
\begin{tabular}{cc}
\includegraphics[width=.25\linewidth]{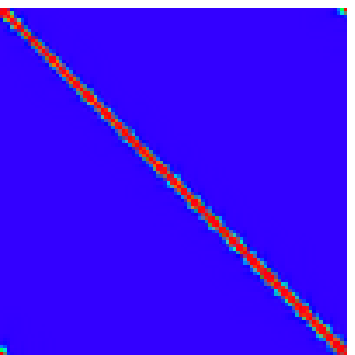} &
\includegraphics[width=.25\linewidth]{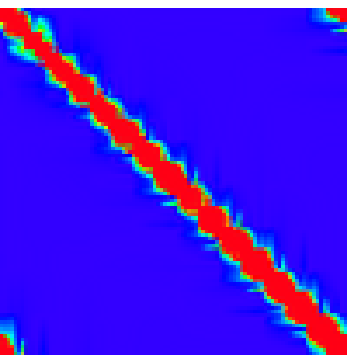} \\
\includegraphics[width=.25\linewidth]{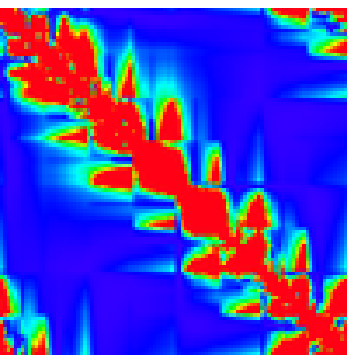} &
\includegraphics[width=.25\linewidth]{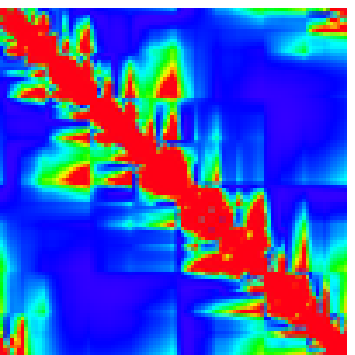} \\
\end{tabular}
\end{center}
\caption{
Density plot of matrix elements $|U_{n,n'}|^2$ for the model (\ref{model}) 
in the computational basis, for 
 $N=2^{12}$,
$k=1$ (top left), $k=10$  (top right), $k=100$ (bottom left) and 
$k=1000$ (bottom right).
Color marks the probability density,
 from blue to red (maximal value).
}
\label{Afig1}       
\end{figure}

\newpage
\section{A2} The spectral analysis of the model (\ref{model}) is obtained 
by a numerical diagonalization of the evolution operator $\hat{U}$. Due to
the unitarity of $\hat{U}$, all the eigenvalues $\lambda$ 
are on the unitary circle, $\lambda  = e^{i \omega}$, where $\omega$ are the 
quasi-energies included in the interval $[0,2\pi)$. The typical examples
for the level spacing statistics $P(s)$ for $\omega$ are shown in 
Fig.\ref{Afig2}. It is remarkable that even for large kick strengths
 ({\it e.g.} $k=1000$) the Wigner-Dyson
statistics of the random matrix theory is not achieved. 
We note that for such values
of $k$ in the sawtooth map
all the eigenstates are delocalized and $P(s)$
is given by the Wigner-Dyson distribution \cite{simone,simone1}.
\begin{figure} [h!]
\epsfxsize=3.2in
\epsffile{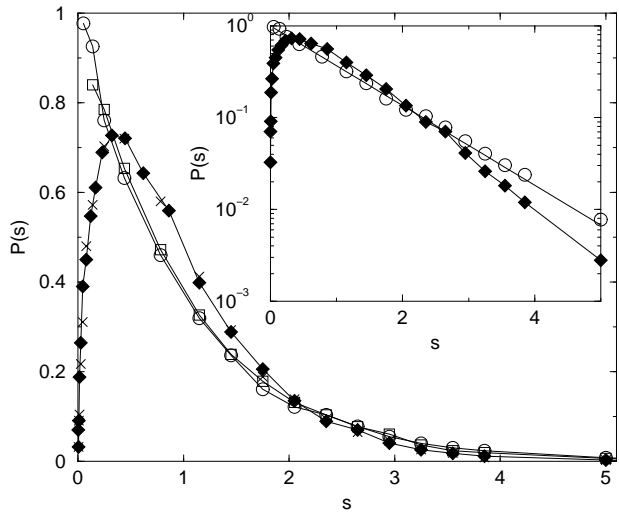}
\caption{Level spacing statistics $P(s)$ for the quasi-energies
of model (\ref{model}), for $n_q=12$ and 
different values of the parameter $k$. 
A transition from the Poisson distribution  
$P(s) =e^{-s}$ at small $k$ values to a distribution which shows the
level repulsion for small $s$ is observed by increasing $k$. 
Data are shown for $k=0.1$ (squares),  
$k=1$ (circles), $k=10$ (diamonds) and $k=1000$ ($\times$'s). 
The inset displays the data for $k=1$ and $k=10$ in a semilogarithmic scale. 
The full line is the Poisson distribution.
}
\label{Afig2}
\end{figure}

\newpage
\section{A3} Our quantum circuit is based on the scheme described in 
\cite{williams}.
We implemented the Pyramidal Algorithm (PYA) for the $D^{(4)}$ wavelet 
transform.
It is based on repeated applications of the operator $D_{2^n}^{(4)}$ 
(the wavelet kernel) and the permutation operator $\Pi_{2^n}$. Here the kernel 
$D_{2^n}^{(4)}$ is the Daubeschies $D^{(4)}$ matrix of size $2^n \times 2^n$.
The operator $\Pi_{2^n}$ realizes the shuffling step on vectors 
$\{v_j\}_{j=1,2^n}$ of 
size $2^n$. The action of $\Pi_{2^n}$ can be regarded as a (classical) 
permutation of the index for the vector $\{v_j\}_{j=1,2^n}$.
The binary representation of index $j$,  $(a_0, a_1, \ldots, a_{n-1})$,
 is mapped into
$(a_{n-1}, a_0, a_1, \ldots, a_{n-2})$, where $a_0$ is the most 
significant bit. 
The classical operator $D^{(4)}$ can be written as
\begin{equation}
D^{(4)}=(D_{4}^{(4)} \oplus I_{2^{n_q}-4}) 
(\Pi_{8} \oplus I_{2^{n_q}-8}) \ldots 
  (D_{2^i}^{(4)} \oplus I_{2^{n_q}-2^i}) 
(\Pi_{2^{i+1}} \oplus I_{2^{n_q}-2^{i+1}}) 
\ldots \Pi_{2^{n_q}} D_{2^{n_q}}^{(4)} 
\label{pyra}
\end{equation}
where $I_M$ is the identity matrix of size $M\times M$ and $\oplus$ is the 
direct 
sum of operators (see Fig.\ref{Afig3}). 

In  a quantum computation, the action of $\Pi_{2^n}$ on the
element $|j\rangle = |a_0, a_1, \ldots, a_{n-1}\rangle$ of the computational 
basis is $\Pi_{2^n} |a_0, a_1, \ldots,
a_{n-2}, a_{n-1}\rangle = |a_{n-1}, a_{n-2}, \ldots, a_1, a_0\rangle$
and it can be implemented via $n-1$
 quantum swaps, each of them  built by $3$ control-not gates. The direct
sums $\Pi_{2^n} \oplus I_{2^{n_q}-2^n}$ and 
$D_{2^n}^{(4)} \oplus I_{2^{n_q}-2^n}$ 
correspond to  multi-controlled 
operators with $n_q-n$ controlling qubits. 
Both $\Pi_{2^{n}}$ and $D_{2^{n}}^{(4)}$ can be 
implemented by a polynomial sequence of elementary gates 
(Toffoli, Control-Not, and one-qubit rotations), therefore 
the above multi-controlled operators are replaced by the product of 
multi-controlled 
elementary gates. Following the procedure proposed in \cite{vedral}, these
multi-controlled gates were implemented through elementary gates with 
the help of an ancilla qubit. The computational cost of a $l$-controlled
gate is linear in the number of controlling qubits $l$.

The wavelet kernel $D_{2^{n}}^{(4)}$ is decomposed into elementary gates
following the factorization proposed 
in \cite{williams} with slight modifications.
The kernel can be written as
\begin{equation}
D_{2^{n}}^{(4)} = (I_{2^{n-1}} \otimes C_1) P_{2^n} (N \otimes 
I_{2^{n-1}})
(N \otimes I_{2^{n-2}} \oplus I_{2^{n-1}}) \ldots
(N \otimes I_{2} \oplus I_{2^{n}-4}) (N \oplus I_{2^n -2}) P_{2^n} 
(I_{2^{n-1}} \otimes C_0) 
\label{kernel}
\end{equation} 
where $P_{2^n}$ is the full permutation matrix which action on the binary
 representation of vector indexes 
is $P_{2^n} (a_0, a_1, \ldots,
a_{n-2}, a_{n-1}) = (a_{n-1}, a_{n-2}, \ldots, a_1, a_0)$
and  $N$ is the not gate.
Here $C_1$, $C_0$ are $2\times 2$ rotation matrices, which can be expressed
via the Daubechies coefficients $c_0, c_1, c_2, c_3$ by defining
\begin{displaymath}
\tilde{C_0}= 2 \left( \begin{array}{cc}
   c_2 & c_3 \\ 	
   c_3 & -c_2
\end{array} \right)
\hspace{2cm}
\tilde{C_1}= \frac{1}{2} \left( \begin{array}{cc}
   \frac{c_0}{c_3} & 1 \\ 	
   1 & -\frac{c_0}{c_3}
\end{array} \right)
\end{displaymath}

\begin{eqnarray}
C_0= \frac{1}{\sqrt{\det{\tilde{C_0}}}} \tilde{C_0} = 
\left( \begin{array}{cc}
   \sin{\theta_0} & \cos{\theta_0} \\ 	
   \cos{\theta_0}\ & -\sin{\theta_0} 
\end{array} \right)
\hspace{2cm}
C_1= \frac{1}{\sqrt{\det{\tilde{C_1}}}} \tilde{C_1}
=  \left( \begin{array}{cc}
\sin{\theta_1} & \cos{\theta_1} \\ 	
   \cos{\theta_1}\ & -\sin{\theta_1}
\end{array} \right)
\label{mat}
\end{eqnarray}
\noindent
where $\theta_0 = \frac{\pi}{3}$ and $\theta_1=\frac{5}{12} \pi$.
The operator $P_{2^n}$ is  implemented by $O(n)$ swap gates.
We note a slight modification in the equation (\ref{mat}), comparing to 
\cite{williams}.
The quantum circuit corresponding to the wavelet kernel (\ref{kernel}) is 
shown in Fig.\ref{Afig3A}. 
Fig.\ref{Afig3B} clarifies the notations used in 
Figs.\ref{Afig3},\ref{Afig3A}.
The total number of elementary gates needed to
implement the kernel circuit scales as $O(n^2)$, thus leading to a $O(n_q^3)$
total complexity for the QWT. 
For our circuit the number of elementary 
gates was $n_g = 1509, 2974, 5237, 8470, 12821, 18462, 25541 $ for 
$n_q=6, 7, 8, 9, 10, 11, 12$. It is assumed that the elementary gates
act between any two qubits.
\begin{figure} [!t!]
\includegraphics[width=.95\linewidth]{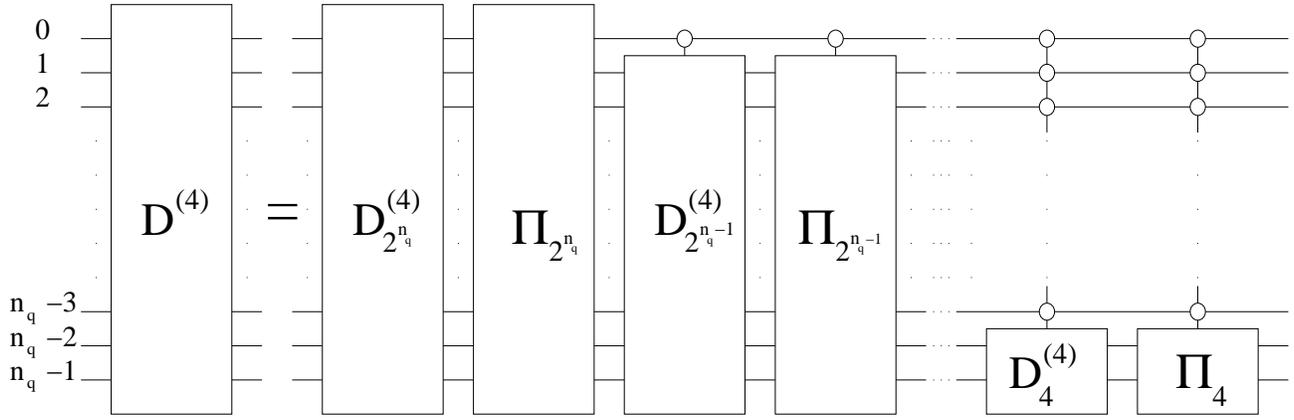}
\caption{
Quantum circuit for the wavelet Transform (\ref{kernel}). 
}
\label{Afig3}
\end{figure}
\begin{figure} [h!]
\includegraphics[width=.95\linewidth]{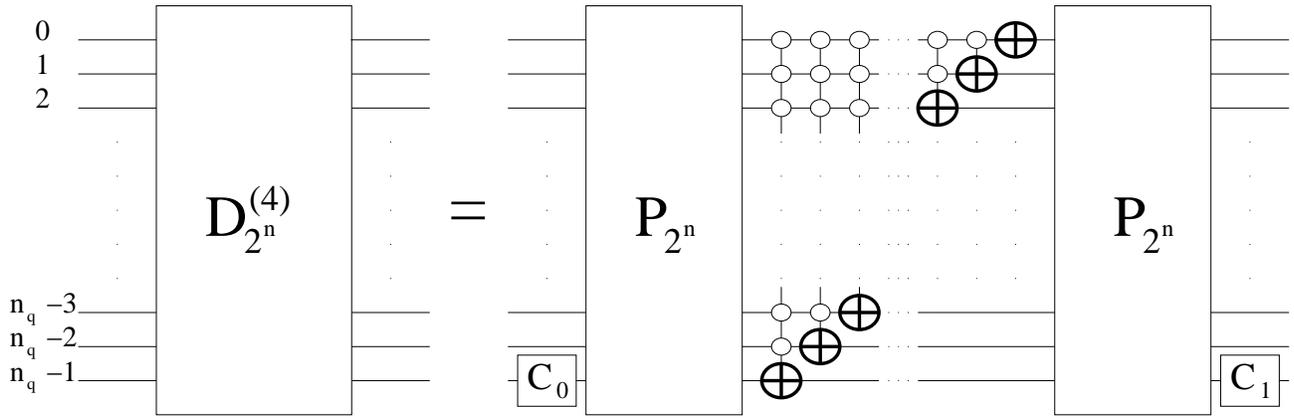}
\caption{
Quantum circuit for the wavelet kernel (\ref{kernel}). {\bf $\bigoplus$} 
represents the Not Operation.
}
\label{Afig3A}
\end{figure}
\begin{displaymath}
\phantom{
\left(\begin{array}{c}
line \\
line \\
line \\
line \\
line \\
line \\
line \\
line \\
line \\
line \\
line \\
line \\
line \\
line \\
line \\
\end{array}\right)
}
\end{displaymath}
\begin{figure} [!h!]
\includegraphics[width=.45\linewidth]{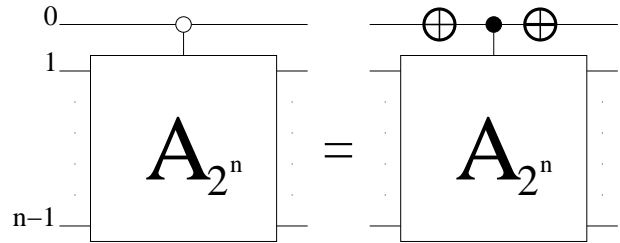}
\caption{
Representation of the $(A_{2^{n-1}} \oplus  I_{2^{n-1}})$ operator, 
{\bf $\bigoplus$} is the Not Operation.
}
\label{Afig3B}
\end{figure}
\begin{displaymath}
\phantom{
\left(\begin{array}{c}
line \\
line \\
line \\
line \\
line \\
line \\
line \\
line \\
line \\
line \\
line \\
\end{array}\right)
}
\end{displaymath}
\newpage

\section{A4}
Here we show the probability distribution in the computational basis for a 
large value of kick strength ($k=1000$) at two different moments of time 
($t=1000$ and $t=10000$) 
(Figs.\ref{Afig4A},\ref{Afig4B}). 
We remark two main features: the distributions have
pronounced peaked structure and the peaks are spread all over the computational
basis. For $t=1000$ the effects of noisy errors are weak so 
that exact and noisy distributions are close (top vs. middle) while
the distribution for static imperfections is already strongly modified
(top vs. bottom).
\begin{figure}[!hb!] 
\epsfxsize=3.2in
\epsffile{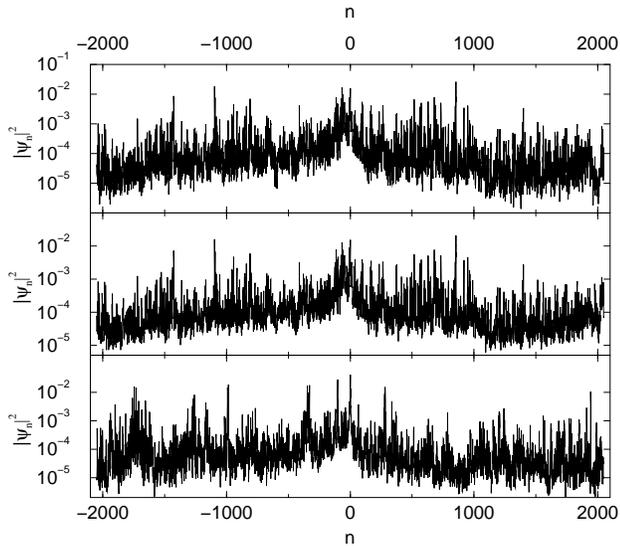}
\caption{
Probability distribution in the computational basis for $n_q=12$, $k=1000$ and
$t=1000$: quantum computation with exact gates (top),
with noisy gates  at $\epsilon=5 \times 10^{-4}$ (middle) 
and static imperfections at $\epsilon= 10^{-4}, \mu=0$ (bottom). 
}
\label{Afig4A}  
\end{figure}
\begin{figure}[!bb!]
\epsfxsize=3.2in
\epsffile{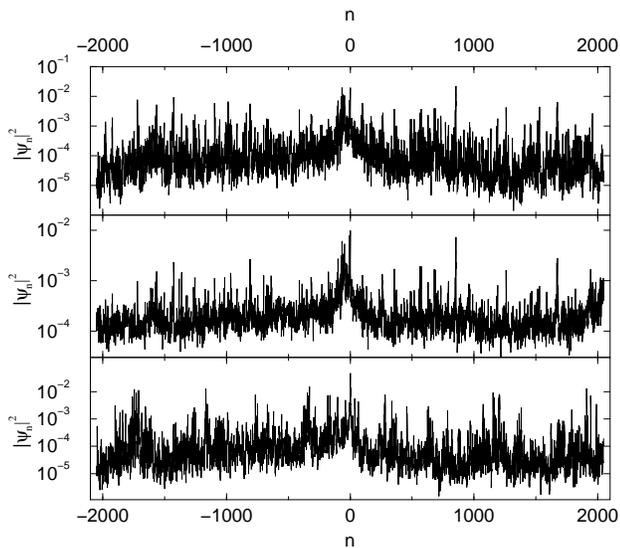}
\caption{Same as Fig.\ref{Afig4A} for $t=10000$.
}
\label{Afig4B}       
\end{figure}

\newpage
\section{A5}
We have also considered another model of static imperfections.
It is obtained from the model of noisy gates by repeating the
same sequence of errors for each application of the evolution
operator $\hat{U}$ in (\ref{model}). 
As in \cite{cat} each noisy gate transformation is  obtained by
diagonalization of nondiagonal part and then by multiplication of 
each eigenvalue by a random phase $\exp(i \eta)$ with 
$-\epsilon/2 <\eta<\epsilon/2$.
This pseudo-static imperfections model is intermediate 
between the two cases considered in the text. The behaviour
of fidelity $f(t)$ is similar to the one shown in Fig.\ref{fig5}a
(see Fig.\ref{Afig5} (top)).
For large and moderate $\epsilon$ the total number of gates $N_g$
is not very large compared to $n_g$ and correlations between different
map iterations can be neglected. Then the gates look like
quasi-random and  the scaling is given by the
relation (\ref{tf1}) with $C \approx 5$. However, 
in the limit of small $\epsilon$ the coherent rotations become dominant
and the data give the scaling $N_g \propto 1/\epsilon$ (see Fig.\ref{Afig5}).
This confirms the generic scaling typical of static imperfections.

\begin{figure}[h!]  
\epsfxsize=3.2in
\epsffile{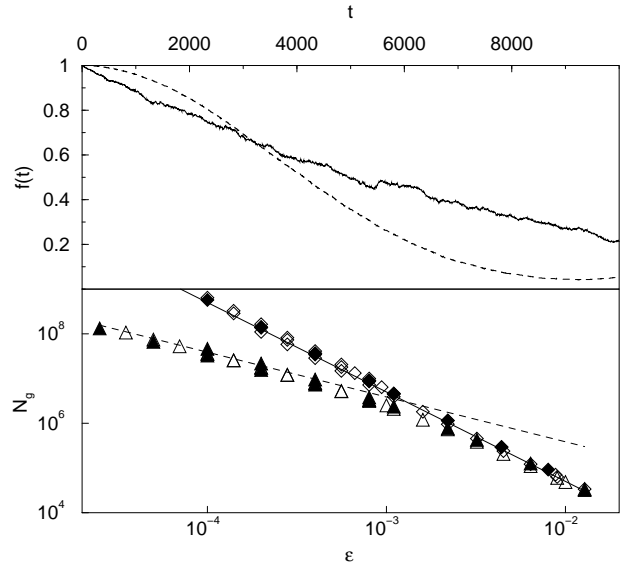}
\caption{Comparison between
noisy gates model and pseudo-static imperfections model. Top panel
shows the behaviour of the fidelity $f(t)$ for $n_q=12$, $k=1, T=1.4$
and $\epsilon = 5 \times 10^{-4}$ (full curve for  noisy gates model)
and $\epsilon = 10^{-4}$ (dashed curve for pseudo-static imperfections).
Bottom panel: scaling for the total number of gates $N_g$
as a function of $\epsilon$ for noisy gates (diamonds) and
pseudo-static imperfections (triangles). Open (full) symbols
correspond to $k=1$ $(k=1000)$. The full
and dashed  straight lines show the dependences
$N_g \propto 1/\epsilon^2$ and $N_g \propto 1/\epsilon$
for noisy gates and psedo-static imperfections respectively.
}
\label{Afig5}
\end{figure}

\end{widetext}

\end{document}